\documentclass[seceq]{ptptex}

\usepackage{wrapft}

\def\cc{\c{c}}

\title{%
Extended NJL model with eight quark interactions
}

\author{
B. \textsc{Hiller}$^{1}$,  
A. A. \textsc{Osipov}$^{1}$, 
A. H.~\textsc{Blin}$^{1}$, J. da~\textsc{Provid\^encia}$^{1}$ 
}

\inst{
$^1$Departamento de Fisica, Centro de Fisica Computacional, Faculdade de Ci\^encias e Tecnologia, Universidade de Coimbra, P-3004-516 Coimbra, Portugal\\
}

\abst{We review the effect on vacuum properties and magnetic catalysis, of 
adding the most general $0$-spin 8-quark $(8q)$ interactions to the combined 
$SU(3)$ flavor Nambu--Jona-Lasinio (NJL) and 6-quark $(6q)$ 't Hooft 
interaction Lagrangian.}

\begin{document}
\maketitle


Since the 8-quark interaction vertices have been introduced \cite{Osipov:2006b}
in the combined 3-flavor NJL and 't Hooft determinantal 
Lagrangians \cite{Nambu:1961}\tocite{Reinhardt:1988} to resolve the instability
of the vacuum associated with the $6q$ terms \cite{Osipov:2006a}, a series of 
phenomenoligical implications for the hadron phenomenology has been studied. 
From the two types of $8q$ interactions considered below, the first 
$(\sim g_1)$ violates the OZI rule, and induces the most relevant effects. 
\begin{eqnarray}   
  && {\cal L}_{8q}^{(1)}\! =\!  
   8\, g_1 
   \left[ (\bar q_iP_Rq_m)(\bar q_mP_Lq_i) \right]^2 
   = \frac{g_1}{32}
   \left[ \mbox{tr} (S-iP)(S+iP)\right]^2
   \! =\!  \frac{g_1}{8}
   \left( S_a^2 + P_a^2\right)^2, \nonumber \\
   &&{\cal L}_{8q}^{(2)}\! =\! 
   \frac{g_2}{16}
   \mbox{tr} \left[(S-iP)(S+iP)(S-iP)(S+iP)\right].
\end{eqnarray}
The trace is taken over the flavor indices $i,j =1,2,3$ of $S_{ij}=S_a 
(\lambda_a)_{ij}=2\bar q_jq_i, P_{ij}=P_a (\lambda_a)_{ij}=2\bar q_j (i\gamma_5) 
q_i$, and $\lambda_a$ denote the Gell-Mann matrices for $a=1,...,8$ and the 
flavor singlet $\lambda_0$. After bosonization of the theory in stationary 
phase approximation one obtains\cite{Osipov:2006b} a set of 3-coupled equations
for the quark condensates $\langle\bar q_i q_i\rangle =h_i/2$ 
\begin{equation}
\label{saddle-1}
   G h_i + M_i - m_i +\displaystyle\frac{\kappa}{16}
   \sum_{j\ne k\ne i} h_jh_k
   +\displaystyle \frac{ g_1}{4}\ 
   h_i \sum_{j=u,d,s}h_j^2 + \displaystyle
   \frac{ g_2 }{2}\ h_i^3=0, 
\end{equation}
where $M_i, m_i$ denote the constituent and current quark masses 
respectively, $G$ is the $4q$ and $\kappa$ the $6q$ coupling strengths. These 
equations must be solved self-consistently with the gap equations
  $ h_i(M_i) + \frac{N_cM_i}{2\pi^2}J_0(M_i^2)=0$
with $J_0$ denoting the quark one loop tadpole. Being of cubic order equations 
(\ref{saddle-1}) can be chosen to have a single real root within a constrained 
set of values for the coupling constants:
$g_1>0, \, g_1 +3g_2>0, \, G>\frac{1}{g_1}\left(\frac{\kappa}{16}\right)^2$.
The resulting effective potential is bounded from below.
The stability constraints allow for a domain of sufficiently small values of 
the $g_1,g_2$ couplings, together with an overall good fit of the low lying 
meson nonet characteristics \cite{Osipov:2007a}.  
Furthermore, noting that the combination  $\xi= G + g_1 (h_u^2+h_d^2+h_s^2)/4$ 
multiplying $h_i$ in (\ref{saddle-1}) also appears in all but the singlet-octet
mixing scalar meson mass expressions \cite{Hiller:2010,Osipov:2007a} one infers
that an increase in the strength of the $8q$ terms can be tuned to a decrease 
in the $4q$ strength $G$, leaving the spectra almost unaffected. In fact 
noticeable changes occur only in the $\sigma$-meson mass, which decreases with 
increasing $g_1$ \cite{Osipov:2007a}. 
One sees also that $g_1$ will steer the value of the effective coupling $\xi$ 
due to any change in the condensates caused by external parameters such as the 
temperature, the chemical potential, the electromagnetic field, etc. It is 
exactly in these cases that one expects the $8q$ interactions to unfold all 
their potential and most importantly, without affecting the vacuum results. As 
an illustration we consider the quark fields minimally coupled to a constant 
magnetic field $H$. In $(2+1)$ and $(3+1)$ D a constant magnetic field $H$ 
catalyzes the dynamical symmetry breaking of the NJL model with $4q$ 
intercations generating a fermion mass even as $G\rightarrow 0$ and the 
symmetry is not restored at any arbitrarily large $H$ \cite{Klevansky:1989}. 
We obtain that the inclusion of higher order multiquark interactions, in 
particular the $8q$ ones, leads to new effects \cite{Osipov:2007b}. In the 
Landau gauge and in the chiral limit the gap equation in presence of $H$ reads 
\begin{equation}
\label{gapH}
-\frac{2  \pi^2 h(M)}{\Lambda^2 N_c}=  f(M^2;\Lambda,|QH|),
\end{equation}
where h(M) is determined through (\ref{saddle-1}).
The couplings $G, \kappa, g_1, g_2$ are contained only in the l.h.s. of 
(\ref{gapH}), and $H$ is present only in $f(M^2;\Lambda,|QH|)$ (see 
\cite{Osipov:2007b} for explicit form of $f$). $Q$ stands for mean quark charge
and $\Lambda$ is the cutoff. 
The function $f$ is singular as $M\rightarrow 0$ for $H\ne 0$. For 
$\kappa=g_1=g_2=0$ the l.h.s. reduces to the constant $\frac{2\pi^2}{\Lambda^2 
N_c G}$, after dividing by $M$, thus eliminating the trivial solution at $M=0$. The gap equation has 
always a nontrivial solution even in the subcritical regime ($G$ not strong enough to break chiral symmetry when $H=0$). The higher multiquark interactions however distort the l.h.s. which cause the order parameter to increase sharply (a secondary magnetic catalysis) with increasing strength of the field at the characteristic scale $H\sim 10^{19} G$. A new phase of massive quarks emerges 
and becomes the stable configuration at that scale \cite{Osipov:2007b}. 
Recently the implications of $8q$ interactions in a strong magnetic background have been considered in an extended $SU(2)$ flavor PNJL model  to study confinement and chiral symmetry restoration, and in the dressed Polyakov loop analysis of the phase diagram of hot quark matter \cite{Gatto}.
\vspace{0.2cm}

\small{{\bf Acknowledgements} We are very grateful to the organizers of ISMD2011, in particular to Prof. A. Nakamura for his interest in our work, exchange of correspondence and discussions. We thank Prof. W. Florkowski from the Intern. Advis. Comm. for his attention to the subjects here presented. Work supported by Funda\cc \~ao para a Ci\^encia e Tecnologia,
SFRH/BPD/63070/2009, CERN/FP/116334/2010 and Centro de F\'isica Computacional, unit 405.}


\begin{thebibliography}{99}

\bibitem{Osipov:2006b} A. A. Osipov, B. Hiller, J. da Providencia, Phys. Lett. B\ \textbf{634} (2006), 48. 
\bibitem{Nambu:1961} Y. Nambu and G. Jona-Lasinio, Phys. Rev. {\bf 122}, 
    345 (1961); {\bf 124}, 246 (1961); 
\bibitem{Hooft:1976} G. 't Hooft, Phys. Rev. D {\bf 14}, 3432 (1976);
    G. 't Hooft, Phys. Rev. D {\bf 18}, 2199 (1978).
\bibitem{Bernard:1988} V. Bernard, R. L. Jaf\mbox{}fe and
    U.-G. Meissner, Phys. Lett. B {\bf 198}, 92 (1987);   
\bibitem{Reinhardt:1988} H. Reinhardt and R. Alkofer, Phys. Lett. B 
    {\bf 207}, 482 (1988).
\bibitem{Osipov:2006a} 
   A. A. Osipov,  B. Hiller, V. Bernard , A. H. Blin, Annals of Phys.  \ \textbf{321} (2006), 2504. 
\bibitem{Osipov:2007a} A. A. Osipov, B. Hiller, A. H. Blin, J. da Providencia, Annals of Phys. \ \textbf{322} (2007), 2021.
\bibitem{Hiller:2010}  B. Hiller, J. Moreira, A.A. Osipov, A.H. Blin, Phys. Rev. D \ \textbf{81} (2010), 116005. 
\bibitem{Klevansky:1989} P. Klevansky and R. H. Lemmer, Phys. Rev.
      {\bf 39}, 3478 (1989). K. G. Klimenko, Theor. Math. Phys. {\bf 89},
      211 (1991). I. V. Krive and S. A. Naftulin, Sov. J. Nucl.
      Phys. {\bf 54}, 897 (1991). V. P. Gusynin, V. A. Miransky, I. A. Shovkovy,
      PRL {\bf 73}, 3499 (1994). 
\bibitem{Osipov:2007b} A. A. Osipov, B. Hiller, A.H. Blin, J. da Providencia, Phys. Lett. B \ \textbf{650} (2007), 262;
B. Hiller, A. A. Osipov, A.H. Blin, J. da Providencia, SIGMA \ \textbf{4} (2008), 024.
\bibitem{Gatto} R. Gatto and M. Ruggieri, Phys. Rev. D  \ \textbf{83} (2011), 034016, R. Gatto and M. Ruggieri, Phys. Rev. D  \ \textbf{82} (2010), 054027.
\end{thebibliography}
\end{document}